# Titanium Contacts to MoS$_2$ with Interfacial Oxide: Interface Chemistry and Thermal Transport

Keren M. Freedy[1], David H. Olson[2], Patrick E. Hopkins[1,2,3], Stephen J. McDonnell[1]

1) Department of Mechanical and Aerospace Engineering, University of Virginia, Charlottesville, VA 22904, USA
2) Department of Materials Science and Engineering, University of Virginia, Charlottesville, VA 22904, USA
3) Department of Physics, University of Virginia, Charlottesville, VA 22904, USA

**ABSTRACT**

The deposition of a thin oxide layer at metal/semiconductor interfaces has been previously reported as a means of reducing contact resistance in 2D electronics. Using X-ray photoelectron spectroscopy with *in-situ* Ti deposition, we fabricate Au/Ti/TiO$_x$/MoS$_2$ samples as well as Au/Ti/MoS$_2$ and Au/TiO$_x$/MoS$_2$ for comparison. Elemental titanium reacts strongly with MoS$_2$ whereas no interface reactions are observed in the two types of samples containing TiO$_x$/MoS$_2$ interfaces. Using time domain thermoreflectance for the measurement of thermal boundary conductance, we find that samples contacted with Ti and a thin TiO$_x$ layer at the interface (≤1.5 nm) exhibit the same behavior as samples contacted solely with pure Ti. The Au/TiO$_x$/MoS$_2$ samples exhibit ~20% lower thermal boundary conductance, despite having the same MoS$_2$ interface chemistry as the samples with thin oxide at the Ti/MoS$_2$ interface. We identify the mechanism for this phenomenon, attributing it to the different interfaces with the top Au contact. Our work demonstrates that the use of thin interfacial oxide layers to reduce electrical contact resistance does not compromise heat flow in 2D electronic devices. We note that the thicknesses of the Ti and TiO$_x$ layers must be considered for optimal thermal transport.

**TEXT**

Contact resistance presents a major obstacle to the success of 2D electronics [1,2]. One approach to the reduction of electrical contact resistance is the deposition of thin oxide interlayers at the interface between the semiconductor and the metal [3-9]. Park *et al.* [3,4] have shown a reduction in electrical contact resistance and an improvement in device stability in $MoS_2$ field effect transistors with 2 nm $TiO_2$ films deposited by atomic layer deposition (ALD) at the interface between the $MoS_2$ and the Ti contact. The observed improvements were attributed to Fermi level de-pinning and interface dipole effects. Kim *et al.* [5], who observed a decrease in Fermi level pinning with 1 nm interfacial $TiO_2$, suggested that the presence of an interfacial oxide reduces the density of metal induced gap states by blocking the penetration of the electron wave function from the metal. Similarly, Kaushik *et al.* [6] concluded from density functional theory that the Schottky barrier height is reduced due to charge-transfer doping from the $TiO_2$ layer to $MoS_2$. They have experimentally shown a twenty four-fold reduction in contact resistance and tenfold improvement in on-current and field effect mobility.

While the use of an interfacial oxide has been found to be highly beneficial to electronic properties in the aforementioned studies, thermal characterization of this interface is relatively lacking. An understanding of thermal transport is crucial as thermal resistances at the contact interface can inhibit heat removal from 2D electronic devices, compromising their performance and reliability [10]. The $MoS_2/SiO_2$ substrate interface present in most 2D devices is typically low, ~14 MW $m^{-2}$ $K^{-1}$[11]. Therefore caution must be taken when introducing additional interfaces to the device that could potentially increase the total resistance of the system. Our previous work has shown that transport across contact interfaces is highly sensitive to the oxide composition of Ti for graphene as well as 3D substrates [12,13]. McDonnell *et al.* [14] have demonstrated that Ti/$MoS_2$ and $TiO_2$/$MoS_2$ interfaces exhibit vastly different chemical compositions and suggested potential detrimental effects on thermal transport due to the higher thermal resistance of $TiO_2$ compared to metallic Ti. They noted that work by Duda *et al.* [15] concludes that the removal of native oxide along with the deposition of a Ti adhesion layer has been found to be critical to lowering thermal resistances at metal-semiconductor interfaces. Similarly, Hopkins *et al.* [16] have shown a substantial decrease in thermal boundary conductance due to the presence of native oxides at metal-semiconductor interfaces. Density functional theory calculations conclude that phonon-phonon coupling and phonon transmission across the metal/$MoS_2$ interface is strongly dependent on the degree of orbital hybridization at the contact, and that stronger chemical and electronic interactions at the contact result in higher thermal boundary conductance [17,18]. This would imply that the inclusion of oxide instead of metal at the $MoS_2$ interface could potentially result in diminished thermal transport, warranting an investigation of the thermal boundary conductance across Ti contacts to $MoS_2$ with interfacial oxides, and this thermal boundary conductance's potential dependence on interface chemistry.

We report a process for electron-beam deposition of an interfacial Ti oxide layer in ultra-high vacuum (UHV) using a partial pressure of $O_2$. The process allows for *in-situ* chemical characterization of the interface with X-ray photoelectron spectroscopy (XPS). We compare the effects of metal (Ti), oxide ($TiO_x$), and metal/oxide heterostructure (Ti/$TiO_x$) films, at a range of thicknesses, deposited on bulk geological $MoS_2$ crystals that are typically used for device fabrication. We use time-domain thermoreflectance (TDTR), an optical pump probe technique, for the measurement of thermal boundary conductance across these interfaces [19-21]. By measuring bulk geological crystals we are able to bypass thermal resistances from substrate/MoS$_2$ interfaces and solely characterize the contact interfaces associated with the MoS$_2$ surface.

Prior to loading to UHV, bulk MoS$_2$ geological crystals (purchased from SPI [22]) were exfoliated with scotch tape to clean the surface by removing the top layer [14]. Preliminary XPS was collected in our ScientaOmicron UHV system [23]. All XPS data were acquired at a pass energy of 50 eV, using an Al K$\alpha$ source with a photon energy of 1486.7 eV. The Mantis QUAD-EVC 4 pocket evaporator was used to deposit Ti onto the sample *in-situ*. The titanium was evaporated at a rate of approximately 1 Å/min. For oxide deposition, a pressure of $5\times10^{-6}$ mbar of ultra-high purity O$_2$ was maintained in the chamber during deposition. XPS was acquired after each Ti and TiO$_x$ deposition. The thicknesses of the deposited layers were calculated using the attenuation of the Mo 3*d* core level intensity via methods described in Supplemental Material. The samples were then capped with 1-2 nm Au in-situ to prevent oxidation in air upon removal from UHV. An additional ~70 nm of Au was deposited in an *ex-situ* e-beam evaporator for TDTR measurements.

TDTR is an optical-pump probe technique that is widely used to characterize interfacial conductance at a variety of metal contacts. An 80 MHz repetition rate laser centered at 800 nm is split into high-power pump and low-power probe paths. The pump is amplitude-modulated using an electro-optic modulator, and frequency-doubled to 400 nm before being focused on the sample surface. The probe is mechanically delayed in time, and monitors the thermoreflectance at the sample surface due to temperature perturbations induced by the pump. We specifically modulate the pump at 10.28 MHz to ensure one-dimensionality in our analysis and minimize sensitivity to potential in-plane transport in the MoS$_2$. Indeed, MoS$_2$ in its few layer and bulk forms have been shown experimentally to have an anisotropic thermal conductivity [24-27]. Modulating at this frequency also improves our sensitivity to the interfacial conductance at these contacts. More information regarding TDTR and its analyses can be found in the literature [19,21] as well as the Supplemental Material.

Thermal boundary conductance, $h_K$, is plotted as a function of TiO$_x$ thickness in Figure 1. In the limit of zero TiO$_x$ thickness, corresponding to samples with pure Ti metal overlayers (Au/Ti/MoS$_2$), the average $h_K$ value was approximately 21.5 ± 5.6 MW m$^2$ K$^{-1}$. This value is in roughly equivalent to the $h_K$ value of the Au/MoS$_2$ reference sample (20.8 ± 1.1 MW m$^2$ K$^{-1}$.). The results are also consistent with previously measured values of metal/MoS$_2$ interfaces [24,28]. We note that the three Au/Ti/MoS$_2$ samples had Ti thicknesses ranging from 2.9 to 5.2 nm and Ti metal thickness had no effect on $h_K$ for these samples, suggesting that the intrinsic resistance of the Ti does not contribute to the overall resistance of the system. Similarly, the figure shows that TiO$_x$/MoS$_2$ samples with TiO$_x$ thicknesses from 1.7 to 4.6 nm all exhibit roughly the same value of 16.0 ± 2.8 MW m$^2$ K$^{-1}$. The lack of thickness dependence of the Ti/MoS$_2$ and TiO$_x$/MoS$_2$ samples indicates that thermal transport is dominated by interfacial resistances and not by the intrinsic thermal resistance of metal or oxide layers.

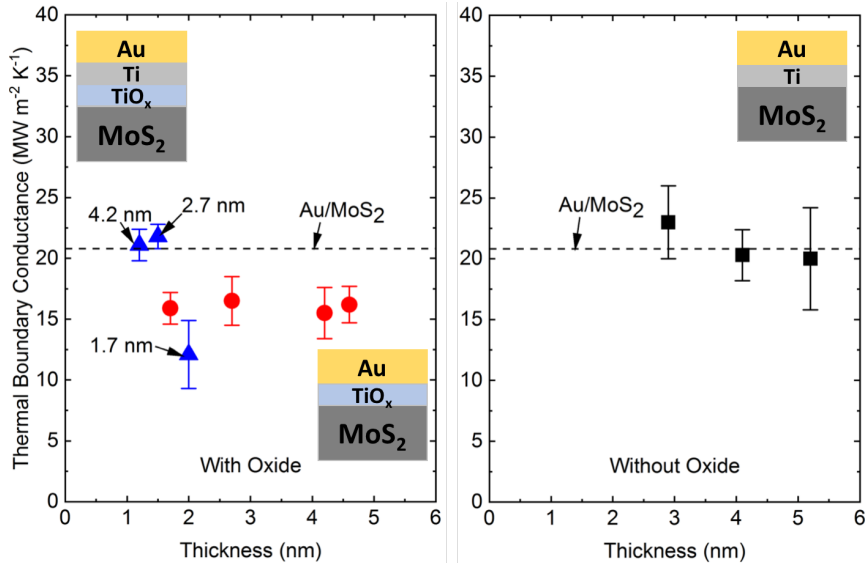

**Figure 1.** Thermal boundary conductance as a function interfacial layer thickness for the MoS$_2$ substrates with Au/Ti (black squares), Au/TiO$_x$ (red circles), and Au/Ti/TiO$_x$ (blue triangles) in addition to a reference sample of Au/MoS$_2$ (dashed line). The arrows indicate the Ti metal thickness for each Ti/TiO$_x$ samples where data is plotted as a function of oxide thickness.

Chemical characterization of the Ti/MoS$_2$ and TiO$_x$/MoS$_2$ interfaces is shown in Figure 2. Spectra of all samples used in this work are included in Supplemental Material. Figure 2(a) shows the core-level XPS spectra before and after the deposition of 4.1 nm Ti metal in UHV. Before the deposition of Ti metal (black curve), MoS$_2$ is characterized by Mo 3$d_{5/2}$ state at ~228.9 eV with a spin orbit splitting value of 3.1 eV, and the S 2$p_{3/2}$ state at ~161.8 eV with a spin orbit splitting of 1.2 eV. Following the deposition of Ti, the spectra exhibit new chemical states including Mo metal (Mo$^0$) at 227.5 eV in the Mo 3$d$ spectrum and Ti-S states in the S 2$p$ spectrum. This result is consistent with previous reports of the deposition of Ti in UHV [14].

Figure 2(b) shows XPS spectra corresponding to a sample with 4.6 nm of $TiO_x$. The spectra indicate that the Mo-S bonds are preserved and no chemical reaction occurs between Ti and the substrate. As previously reported by McDonnell *et al.* [14], the presence of a partial pressure of oxygen during the deposition of Ti on $MoS_2$ inhibits the reaction between them as Ti reacts with oxygen impinging on the surface of the substrate during deposition. The Mo $3d$ and S $2p$ core levels exhibit a 0.64 eV shift to higher binding energy, corresponding to a change in the position of the Fermi level. This indicates that the presence of an oxide overlayer causes n-type doping in the sample. This result is similar to that of Kaushik *et al.* [6] who reported a 0.5 eV core-level shift for a 2 nm of ALD $TiO_2$ on $MoS_2$. We note that the oxide which forms under the deposition conditions in our UHV chamber is comprised of two chemical states. The $TiO_2$ state has its $2p_{3/2}$ component at 459.2 eV with a spin-orbit splitting of 5.7 eV and comprises ~80% of the oxide layer deposited. The second chemical state, which appears at 457.65 eV with a spin-orbit splitting of 5.5 eV corresponds to $Ti_2O_3$ [29,30]. Spectral deconvolution of the oxide is shown in Supplemental Material.

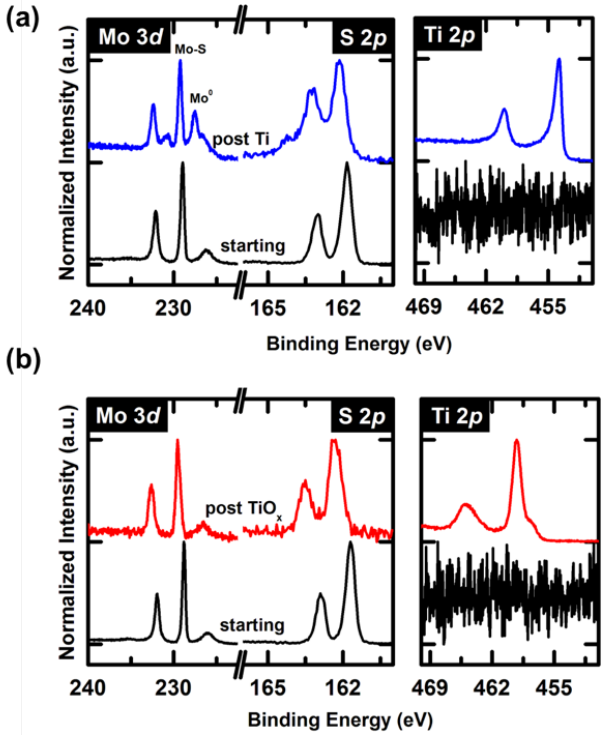

**Figure 2.** Core level XPS spectra before and after the deposition of (a) 4.1 nm Ti on MoS$_2$ and (b) 4.6 nm TiO$_x$ on MoS$_2$

Unlike the pure metal and oxide samples, $h_K$ of the Au/Ti/TiO$_x$/MoS$_2$ samples exhibits a decrease with increasing oxide thickness, with no apparent dependence on metallic Ti thickness, as shown in Figure 1. The two samples with TiO$_x$ thicknesses ≤ 1.5 nm have $h_K$ values comparable to that of the Au/Ti/MoS$_2$ and Au/MoS$_2$ samples, whereas the Au/Ti(1.7 nm)/TiO$_x$(2.0 nm)/MoS$_2$ is comparable to the Au/TiO$_x$/MoS$_2$ samples with no metal overlayer, within uncertainty. The reduction in $h_K$ for this sample could be due to two possible reasons. The first is the increase in the oxide thickness compared to the heterostructures with TiO$_x$ thickness ≤ 1.5 nm. However, given that no oxide thickness dependence for $h_K$ of the Au/TiO$_x$/MoS$_2$ samples is observed, the increased oxide thickness is unlikely to be the dominant factor here. The second explanation is that the thickness of the Ti in the heterostructure (1.7 nm) is quite thin.

From previous works, it has been shown that a reduction in the interfacial conductance is observed as the thickness of an interfacial adhesion layer becomes very thin (e.g., Cu and Cr) [31]. This reduction in our experiment is explained using similar predictions of accumulated thermal boundary conductance, whereby phonons in Ti with wavelengths less than the total Ti thickness participate in transport across the interface. In this way, the reduction in the population of Ti phonons due to a decrease in total Ti thickness ultimately results in a reduced $h_K$ at the interface. We believe this to be the case when the thickness of the Ti becomes very thin in the heterostructures, making our results are consistent with those of Jeong *et al.* [31].

The XPS spectra of a Ti(2.7 nm)/TiO$_x$(1.5 nm)/MoS$_2$ sample are shown in Figure 3. The TiO$_x$/MoS$_2$ interface (red curve) is chemically identical to that shown in Figure 2(b), exhibiting a n-type Fermi level shift and no other chemical changes following the deposition of Ti metal (blue curve). The only observable changes are broadening of the peaks and an increase in noise, which occurs due to scattering and attenuation in the TiO$_x$ and Ti overlayers [32]. The lack of interface reactions with the presence of interfacial oxide is one possible explanation for the Fermi level de-pinning effect reported by others [3-5]. Fermi level pinning has been attributed to interfacial reaction products which create new electronic states within the semiconductor band gap [33]. By blocking interface reactions via direct contact with an unreactive oxide layer, MoS$_2$ retains its intrinsic band structure with no new states which could pin the Fermi level.

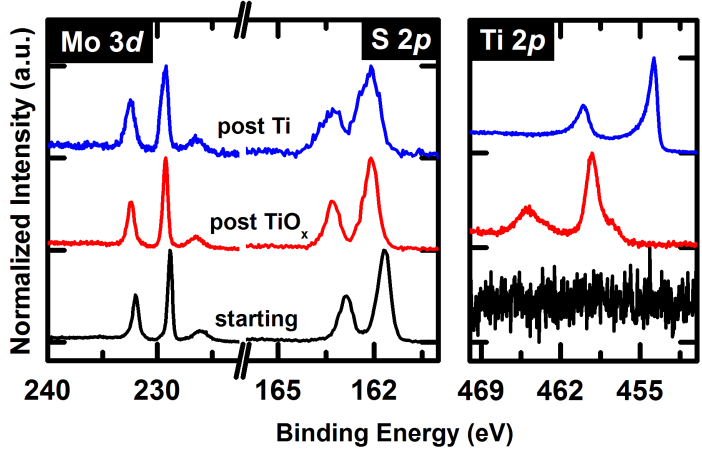

**Figure 3.** Core-level XPS spectra of the Ti/TiO$_x$/MoS$_2$ samples

It is clear from comparison of Figure 2(b) and Figure 3 that interface chemistry does not explain the differences in behavior of $h_K$, as the TiO$_x$/MoS$_2$ interface in both the pure oxide and metal/oxide samples are chemically identical. Therefore, the behavior of $h_K$ is likely dominated by one of the other interfaces present in the device. For the metal/oxide heterostructure sample these interfaces include the Au/Ti and Ti/TiO$_x$, whereas the oxide sample has the Au/TiO$_x$. The resistance from the Au/Ti interface is not a contributing factor, since thermal transport across metal/metal interfaces is governed by electrons near the Fermi energy, yielding thermal boundary conductance values that are far higher than those across metal/non-metal interfaces (i.e., negligible thermal resistances at these metal/metal interfaces) [34-37]. The negligible contribution of this interface is evident in Figure 1, which shows that Au/Ti/MoS$_2$ exhibits the same $h_K$ as the Au/MoS$_2$ reference sample, and consistent with previous measurements of thermal boundary conductance at Au/graphene and Au/Ti/graphene interfaces [38].

To determine the extent of the contribution of the thermal conductances across the Au/TiO$_x$, Au/Ti, and Ti/TiO$_x$ interfaces to the overall measured $h_K$, we fabricate several reference samples. We first deposited 80 nm Au on single crystal sapphire (Al$_2$O$_3$) with a 5 nm Al adhesion layer in a cleanroom evaporator at high vacuum (HV). $Al_2O_3$ was selected as our substrate to maximize sensitivity to potentially large conductances. The conductance at the Au/Al/$Al_2O_3$ interface was found to be 49 ± 5.0 MW m$^{-2}$ K$^{-1}$. Following this measurement, the surface of the Au was cleaned via UV-$O_3$ exposure as described elsewhere [39] to remove adventitious carbon. The sample was loaded back to UHV. Since the process was found to leave residual oxygen on the Au surface, the sample was capped with ~ 2 nm Au in UHV to create a pristine surface before depositing subsequent Ti layers. Three separate samples with Ti (2.3 nm), $TiO_x$ (2 nm), and a Ti (2.2 nm)/$TiO_x$ (4.4 nm) were created and capped with ~1-2 nm Au. To account for the contribution of residual oxygen at the interface between the HV and UHV Au, we fabricate a reference sample that is a UV-$O_3$ treated HV Au sample that is then capped with Au in UHV. This sample is measured to determine the contribution of the oxygenated Au interface to the total interfacial conductance in structures containing Ti and $TiO_x$.

Following the UHV deposition processes, the samples were transferred back to the HV electron-beam evaporator, and capped with ~ 67 nm of Au following an $O_2$ plasma cleaning procedure. TDTR was performed on the samples, fitting for the conductance across the newly deposited interfacial structure as the underlying Au/Al/$Al_2O_3$ interfacial conductance was measured prior to the deposition of the structures. The conductance across the Au/Au interface containing residual oxygen from the UV-$O_3$ process is measured to be 376 ± 31 MW m$^{-2}$ K$^{-1}$, and is accounted for in subsequent derivations of Au/Ti and Au/$TiO_x$ conductances. When accounting for this additional resistance, $h_K$ of the Au/Ti interface is found to be 1680 ± 190 MW m$^{-2}$ K$^{-1}$. We estimate this to be the lower bound for the conductance at the Au/Ti interface, based on the limitations of TDTR to measure ultrahigh boundary conductances, quasi-ballistic influences as a result of the thinness of the Ti layer, and extrinsic or chemical effects that prevent the Au/Ti interface from being an otherwise perfect interface. One possible explanation will be addressed later. Regardless, the conductance of metal-metal interfaces is quite high—the electron diffuse-mismatch model predicts a conductance of 5970 MW m$^{-2}$ K$^{-1}$ at the Au/Ti interface [34]. Thus, the contribution of the Au/Ti interface in the total interfacial resistance is negligible compared to others resistances present in these systems.

The Au/TiO$_x$ (2 nm)/Au conductance is measured to be 44.5 ± 3.2 MW m$^{-2}$ K$^{-1}$, resulting in an interfacial conductance at the Au/TiO$_x$ interface to be 101 ± 11 MW m$^{-2}$ K$^{-1}$. The measured conductance across the Au/Ti (2.2 nm)/TiO$_x$ (4.4 nm)/Au interface is 65.2 ± 4.5 MW m$^{-2}$ K$^{-1}$, and suggests that the conductance of the Ti/TiO$_x$ interface, with a value of 459 ± 87 MW m$^{-2}$ K$^{-1}$, is large compared to that across the Au/TiO$_x$. Our results imply that the Au/TiO$_x$ interface presents a non-negligible thermal resistance due to the relative weak atomic interactions between Au and TiO$_x$; in fact, in this case of this Au/TiO$_x$/Au/Al/Al$_2$O$_3$ multilayer film system, the Au/TiO$_x$ offers the limiting thermal resistance to heat flow. A complete derivation of all of the above values can be found in the Supplemental Material.

The thermal conductance across the Au/TiO$_x$/MoS$_2$ interface can be modeled with a series resistance approach:

$$\frac{1}{h_{K,meas}} = \frac{1}{h_{K,Au/TiO_x}} + \frac{1}{h_{K,TiO_x/MoS_2}},$$

where $h_{K,meas}$ is the measured conductance across the Au/TiO$_x$/MoS$_2$ interface (averaged over all TiO$_x$ thicknesses), $1/h_{K,Au/TiOx}$ is the resistance of the Au/TiO$_x$ interface, and $1/h_{K,TiOx/MoS2}$ is the resistance of the TiO$_x$/MoS$_2$ interface. Again, we neglect the contribution from the intrinsic resistance of the TiO$_x$ layer based on the constant boundary conductance observed for the range of TiO$_x$ thicknesses that are presented in Fig. 1. Taking $h_{K,meas}$ to be 16.0 ± 2.8 MW m$^{-2}$ K$^{-1}$, and $h_{K,Au/TiOx}$ from our measurements on this control interface discussed above, we calculate $h_{K,TiOx/MoS2}$ to be 19.1 ± 3.8 MW m$^{-2}$ K$^{-1}$. The larger comparable conductance at the TiO$_x$/MoS$_2$ interface as compared to $h_{K,meas}$ suggests that the Au/TiO$_x$ interfacial conductance is again playing a non-negligible role in the reduction of heat transport across these interfaces, albeit, the TiO$_x$/MoS$_2$ represents the dominant thermal resistance in this system. This reduction can be circumvented by implementing a Ti/TiO$_x$ heterostructure at the interface, provided that the Ti layer is thicker than ~ 2 nm as discussed previously, whereby we see an increased boundary conductance as compared to just a TiO$_x$ adhesion layer. In all, this also suggests that the Ti/TiO$_x$ conductance is negligible compared to that of the Au/TiO$_x$ and TiO$_x$/MoS$_2$ interface, allowing for a compromise of transport properties from both an electrical and thermal perspective. The values of the various interfaces with Ti and MoS$_2$ are summarized in Table I.

**Table I.** Measured and derived values of Ti/TiO$_x$ structures on MoS$_2$ and sandwiched between Au. The * denotes values that are derived from measurements. The † denotes that the average of the heterostructures with TiO$_x$ thicknesses of 1.5 and 2.0 nm, omitting that with 1.7 nm Ti. Other values are averaged from all samples of that type.

| Interface | Au/Ti* | Au/TiO$_x$* | Ti/TiO$_x$* | Au/MoS$_2$ | Au/Ti/MoS$_2$ | Au/TiO$_x$/MoS$_2$ | Au/Ti/TiO$_x$/MoS$_2$† | TiO$_x$/MoS$_2$* |
|---|---|---|---|---|---|---|---|---|
| $h_K$ (MW m$^{-2}$ K$^{-1}$) | 1680 ± 190 | 101 ± 11 | 459 ± 87 | 20.8 ± 1.1 | 21.1 ± 5.7 | 16.0 ± 2.8 | 21.5 ± 1.6 | 19.1 ± 3.8 |
| $R_k$ (m$^2$ K GW$^{-1}$) | 0.59 ± 0.07 | 9.91 ± 1.07 | 2.18 ± 0.41 | 48.1 ± 2.54 | 47.4 ± 12.8 | 62.5 ± 10.9 | 46.5 ± 3.46 | 52.4 ± 10.4 |

XPS characterization of the Au/Ti and Au/TiO$_x$ interfaces shows that chemical bonding occurs at Au/Ti interfaces while no chemical interactions are observed in Au/TiO$_x$ interfaces. The formation of intermetallic compounds in Au/Ti interfaces deposited in UHV at room temperature has been previously reported by others [40-42]. Figure 4 shows XPS acquired on Ti and TiO$_x$ samples deposited in our UHV system before and after the deposition of ~3 Å of Au. In the Ti 2$p$ spectrum of the Ti metal sample shown in (a), the spectrum exhibits a core level shift of 0.15 eV as well as broadening due to the presence of Au-Ti bonding at the interface following the deposition of Au. The TiO$_x$ spectra shown in (b) acquired before and after Au deposition overlap perfectly showing no change in binding energy or line shape. In the Au 4$f$ spectrum in (c), Au deposited on Ti exhibits a prominent asymmetry and 0.33 eV shift to higher binding energy which is characteristic of intermetallic formation [43], while Au deposited on TiO$_x$ retains the line shape and binding energy of elemental Au indicating no interaction occurs. The observed bonding at the Au/Ti interface is a potential explanation for the lower measured $h_K$ of the Au/Ti interface in comparison with the value calculated based on the electron diffuse-mismatch model, which does not account for the formation of an intermetallic compound at the interface [34]. It has been shown by others that the thermal boundary conductance can become dominated by the thermal conductance of an interfacial compound layer [34]. Furthermore, intermetallic compounds have been found to exhibit low values of thermal conductivity in comparison with their pure metal constituents [44]. Nevertheless, the electron mediated thermal transport at metallic Au/Ti interface results in $h_K$ value that is far higher than that corresponding to Au/TiO$_x$.

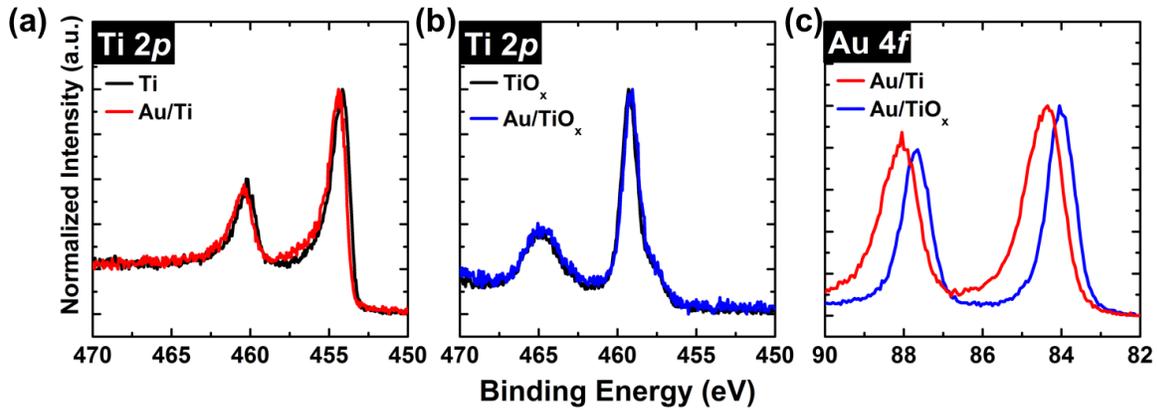

**Figure 4.** Ti 2*p* spectra acquired on (a) Ti and (b) TiO$_x$ films before and after the deposition of Au, with Au 4*f* spectra corresponding to these samples shown in (c).

In summary, we find that Au/TiO$_x$/MoS$_2$ exhibit ~20% lower thermal boundary conductance than Au/Ti/MoS$_2$. Samples with a thin TiO$_x$ layer (~ 1-1.5 nm) at the interface between Ti and MoS$_2$ exhibit the same thermal boundary conductance as those with pure Ti metal. The difference in $h_K$ between the TiO$_x$/MoS$_2$ and Ti/TiO$_x$/MoS$_2$ samples is observed despite the chemically identically TiO$_x$/MoS$_2$ interfaces present in both samples. The differences in $h_K$ arise due to the different interfaces with the top Au contact. Whereas Au/Ti has negligible resistance, that of Au/TiO$_x$ is substantial making this interface the dominant resistor in the system. Our results suggest that thin interfacial oxide layers which can be used to enhance electronic properties have no negative impact on thermal transport in 2D electronic devices. The thickness of the Ti layer in the Ti/TiO$_x$ structure must be considered when implementing this type of contact.

**ACKNOWLEDGEMENTS**

We appreciate funding from the US Department of Defense, Multidisciplinary University Research Initiative through the Army Research Office, Grant No. W911NF-16-1-0406. D. H. Olson is grateful for funding from the National Defense Science and Engineering (NDSEG) and Virginia Space Grant Consortium (VSGC) fellowships.

# Supplemental Material:

# Titanium Contacts to MoS$_2$ with Interfacial Oxide: Interface Chemistry and Thermal Transport

Keren M. Freedy[1], David H. Olson[2], Patrick E. Hopkins[1,2,3], Stephen J. McDonnell[1]

1) Department of Mechanical and Aerospace Engineering, University of Virginia, Charlottesville, VA 22904, USA
2) Department of Materials Science and Engineering, University of Virginia, Charlottesville, VA 22904, USA
3) Department of Physics, University of Virginia, Charlottesville, VA 22904, USA

1. **Time Domain Thermoreflectance**

To determine the boundary conductances across the TiO$_x$, Ti, and heterostructure interfaces, we implement time-domain thermoreflectance. In our approach, 100 fs pulses emanating from an 80 MHz laser centered at 800 nm are split into a high-energy pump path and a low-energy probe path. The pump is electro-optically modulated and frequency doubled to 400 nm, and creates a frequency-dependent heating event at the sample surface. The probe path is mechanically delayed in time, and monitors the thermoreflectance at the samples surfaces as a function of delay time. This provides the temporal resolution to generate the cooling curve at the sample surface, to which we fit using our radially symmetric heat diffusion equation. The pump and probe 1/e$^2$ radii are ~8 and ~4.5 mm, respectively, after being focused through a 10 x objective. We specifically fit for the conductance across the Au/MoS$_2$ interface, leaving the intermediate structures and their resistances as one. The volumetric heat capacity of the Au capping layer is assumed to be 2.49 MJ m$^{-3}$ K$^{-1}$, while the thickness is confirmed with a reference sample during the deposition procedure. We also assume a thermal conductivity of 150 W m$^{-1}$ K$^{-1}$ for the 72 nm Au capping layer, based on previous measurements, although we find that we are not terribly sensitive to the value at these spot sizes. The volumetric heat capacity of MoS$_2$ is taken from the literature as 1.89 MJ m$^{-3}$ K$^{-1}$.[1] Sensitivity to the above parameters is shown in Figure S1 at a modulation frequency of 10.28 MHz, where we follow the formalism of Gundrum *et al.* [2] to determine sensitivity to our parameters. The cross- and in-plane thermal conductivities of MoS$_2$ are chosen to be 2 and 100 W m$^{-1}$ K$^{-1}$, respectively [3,4], while the Au/MoS$_2$ interfacial conductance, which includes the Ti or TiO$_x$ layer, is chosen to be 20 MW m$^{-2}$ K$^{-1}$.

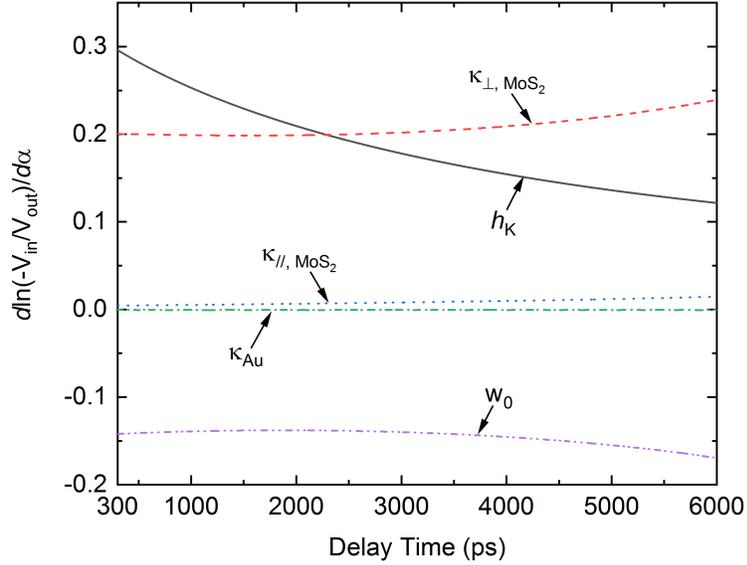

**Figure S1.** Sensitivity, $d\ln(-V_{in}/V_{out})/da$ of relevant parameters in the experiment. Sensitivities to the cross- and in-plane thermal conductivities of the MoS$_2$ are denoted by $k_{\perp, MoS2}$ and $k_{//, MoS2}$, respectively. Also shown in the figure are sensitivities to the thermal conductivity of the Au capping layer, $k_{Au}$, the boundary conductance at the Au/MoS$_2$ interface, $h_K$, and the spot size, $w_0$.

## 2. Derivation of Thermal Conductances

We model each of the thermal conductances measured as series resistors models, accounting for the interfaces present in the system, and disregarding the intrinsic resistances of the layers due to the constant thermal boundary conductances observed in our samples as a function of thickness. For completeness, we have placed the tables from the main manuscript in this document, shown in Table I. The Au/Ti (2.3 nm)/Au/O/Au interfacial region can be modeled as such:

$$\frac{1}{h_{K,Au/Ti/Au/O/Au}} = \frac{2}{h_{K,Au/Ti}} + \frac{1}{h_{K,Au/O/Au}}.$$

In this equation, $h_{K,Au/Ti/Au/O/Au}$ represents the total, measured conductance, $h_{K,Au/Ti}$ represents the conductance at the Au/Ti interface, and $h_{K,Au/O/Au}$ accounts for the resistance of the Au/Au oxygenated layer. Because we have previously measured $h_{K,Au/O/Au}$, we can calculate $h_{K,Au/Ti}$ to be $1680 \pm 190$ MW m$^{-2}$ K$^{-1}$. In a similar manner, the conductance at the Au/TiO$_x$ (2.0 nm)/Au/O/Au region can be defined as

$$\frac{1}{h_{K,Au/TiOx\,(2.0\,nm)/Au/O/Au}} = \frac{2}{h_{K,Au/TiOx}} + \frac{1}{h_{K,Au/O/Au}},$$

where $G_{Au/TiOx}$ is the conductance across the Au/TiO$_x$ interface. $G_{Au/TiOx}$ can thus be calculated to be $101 \pm 11$ MW m$^{-2}$ K$^{-1}$. Finally, the Ti (2.2 nm)/TiOx (4.4 nm) heterostructure can be modeled under the following equation:

$$\frac{1}{h_{K,Au/Ti\ (2.2\ nm)/TiOx\ (4.4\ nm)/Au/O/Au}} = \frac{1}{h_{K,Au/Ti}} + \frac{1}{h_{K,Ti/TiOx}} + \frac{1}{h_{K,Au/TiOx}} + \frac{1}{h_{K,Au/O/Au}}.$$

In this equation, $h_{K,Ti/TiOx}$ is the conductance at the Ti/TiO$_x$ interface. If one assumes that the Ti/TiO$_x$ interfacial resistance is negligible (i.e., $1/h_{K,Ti/TiOx}$ -> 0), then the calculated conductance taking into account just $h_{K,Au/Ti}$, $h_{K,Au/TiOx}$, and $h_{K,Au/O/Au}$ results in a net conductance across the heterostructure of $76.0 \pm 13.0$ MW m$^{-2}$ K$^{-1}$. This is in good agreement with the measured value of the heterostructure ($65.2 \pm 4.5$ MW m$^{-2}$ K$^{-1}$), and suggests that the interfacial conductance at the Ti/TiO$_x$ interface is relatively high at $459 \pm 87$ MW m$^{-2}$ K$^{-1}$. This same type of analysis, as shown in the manuscript, is used to determine the thermal conductance at the TiO$_x$/MoS$_2$ interface from the values determined above.

**Table S1.** Measured and derived values used in the above analysis

| Interface | $h_K$ (MW m$^{-2}$ K$^{-1}$) | $R_k$ (m$^2$ K GW$^{-1}$) |
|---|---|---|
| Au/O/Au | $376 \pm 31$ | $2.66 \pm 0.22$ |
| Au/Ti (2.3 nm)/Au/O/Au | $260 \pm 20$ | $3.85 \pm 0.29$ |
| Au/TiO$_x$ (2 nm)/Au/O/Au | $44.5 \pm 3.2$ | $22.5 \pm 1.6$ |
| Au/Ti (2.2 nm)/TiO$_x$ (4.4 nm)/Au/O/Au | $65.2 \pm 4.5$ | $15.3 \pm 1.0$ |
| Au/O* | $752 \pm 61$ | $1.33 \pm 0.11$ |
| Au/Ti* | $1680 \pm 190$ | $0.59 \pm 0.07$ |
| Au/TiO$_x$* | $101 \pm 11$ | $9.91 \pm 1.07$ |
| Ti/TiO$_x$* | $459 \pm 87$ | $2.18 \pm 0.41$ |

## 3. Core Level Spectra

Spectra of all samples measured in this work are shown here in Figures S2-4. With thicker overlayers, the Mo 3*d* and S 2*p* signals are increasingly diminished in intensity due to attenuation. In Figure S2, it is apparent that the ratio of the Mo$^0$ to MoS$_2$ peak intensities increases with Ti thickness. This difference in the relative quantities of the chemical states, a result of differences in Ti thickness, has no measurable effect on $h_K$. In Figure S3, which shows spectra corresponding to the Ti/TiO$_x$ interfaces, all samples exhibit identical interface chemistry since no chemical reaction takes place. The same is true for the Ti/TiO$_x$/MoS2 samples in Figure S4.

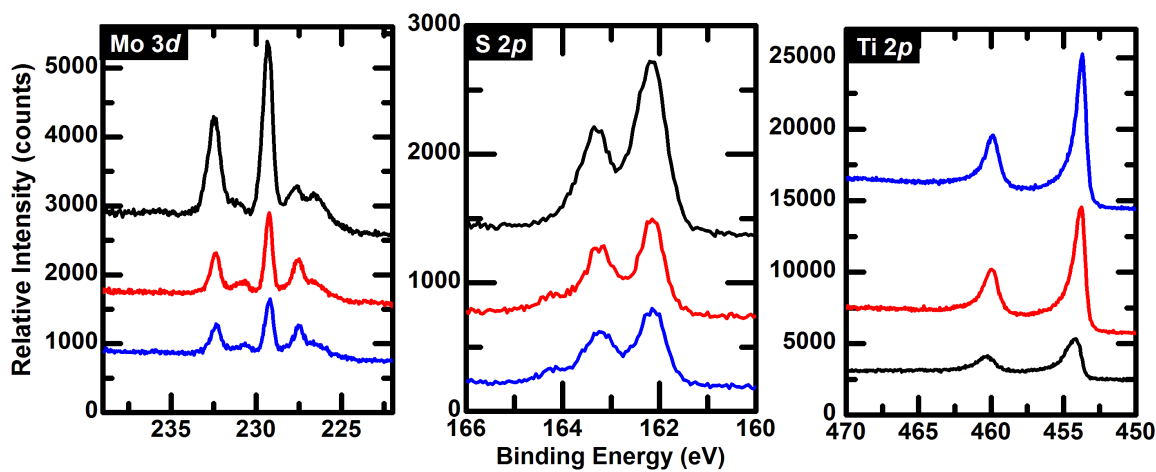

**Figure S2.** XPS spectra of the Mo 3d, S 2p, and Ti 2p core-levels for all Ti/MoS2 samples with Ti thicknesses of 2.9 nm (black), 4.1 nm (red), and 5.2 nm (blue).

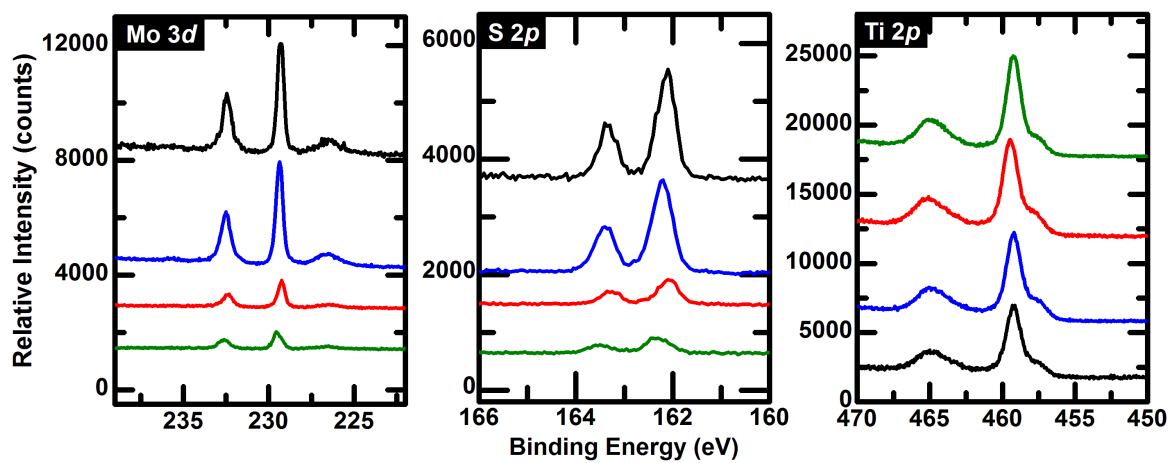

**Figure S3.** XPS spectra of Mo 3d, S 2p, and Ti 2p core levels for TiOx/MoS2 with TiOx thickness of 1.7 nm (black), 2.7 nm (blue), 4.2 nm (red), and 4.6 nm (green)

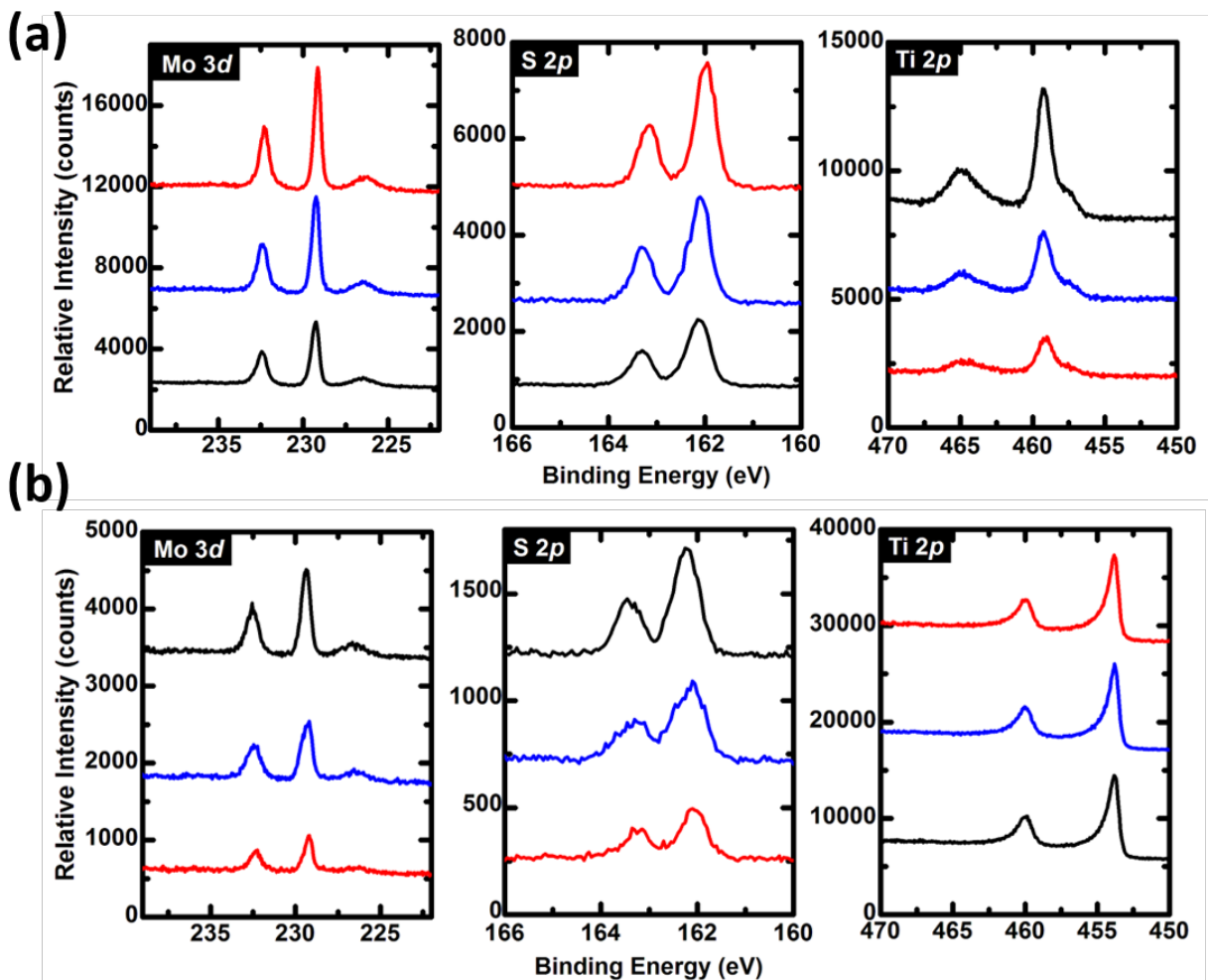

**Figure S4.** XPS spectra of Mo 3d, S 2p, and Ti 2p core levels for Ti/TiOx/MoS2 samples after the deposition of TiOx in (a) and then Ti in (b). The red curve corresponds to 1.2 nm TiOx + 4.2 nm Ti, the blue curve corresponds to 1.5 nm TiOx + 2.7 nm Ti, and the black curve corresponds to 2.0 nm +1.7 nm Ti.

4. **Composition of Ti Oxide**

A representative spectral deconvolution of the oxide is shown in Figure S5. The integrated intensities were determined from fits to the experimental data constructed using XPS fitting software from kolXPD [5]. $TiO_2$ comprises ~80% of the oxide layer in all samples.

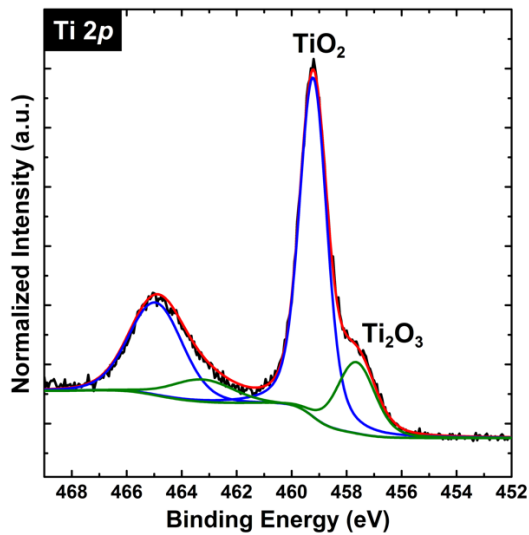

**Figure S5.** Representative spectral deconvolution of the Ti 2p core level for TiO$_x$ deposited in an oxygen partial pressure of 5 ×10$^{-6}$ mbar at a deposition rate of 1 Å/min

## 5. Layer Thickness Calculation

The thickness of the deposited layers was calculated using the integrated intensities of the Mo 3$d$ core level of the MoS$_2$ substrate before and after deposition. Examples of spectra acquired with metal and oxide overlayers are shown in Figure S6.

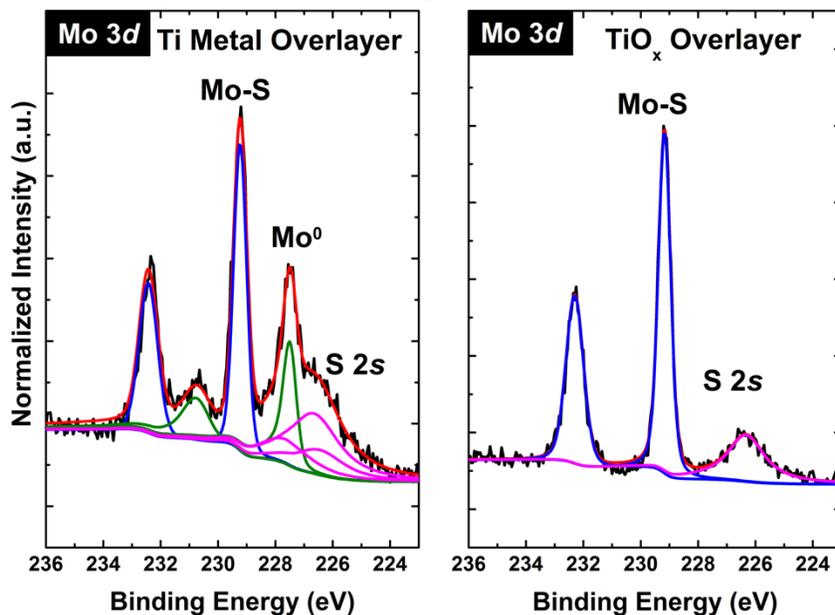

**Figure S6.** Representative spectral deconvolution of the Mo 3$d$ core level with metal and oxide overlayers. In the MoS$_2$ with a Ti metal overlayer on the left, the S 2$s$ curve in the spectrum corresponds to a Mo-S and two Ti-S states.

When a thin film is deposited on an infinite thick substrate, the measured substrate intensity $I$ decays as a function of the film thickness $d$ according to [6-8]

$$I = I_0 e^{-\frac{d}{\cos\theta * EAL}}$$

where $I_0$ is the intensity of the substrate before film deposition, $\theta$ is the photoelectron takeoff angle, and $EAL$ is the effective attenuation length. In our ScientaOmicron system, the take-off angle is 45°. Solving for thickness yields

$$d = -\cos\theta * EAL * \ln\left(\frac{I}{I_0}\right).$$

The $EAL$ of Mo $3d$ in the overlayers determined using the NIST EAL database.[9] For samples with a metallic Ti overlayer, a weighted average of the EAL values of Mo $3d_{5/2}$ in Ti (26.008 Å) and Mo $3d_{5/2}$ in Mo metal (15.638 Å). The percentage of Mo metal assuming a homogeneous overlayer of Ti and Mo.